\documentstyle[12pt,epsf]{article}
\newcommand{\agt}{\,\rlap{\lower 3.5 pt \hbox{$\mathchar \sim$}} \raise 1pt
 \hbox {$>$}\,}
\newcommand{\alt}{\,\rlap{\lower 3.5 pt \hbox{$\mathchar \sim$}} \raise 1pt
 \hbox {$<$}\,}

\begin{document}
\title{\vskip-3cm{\baselineskip14pt
\centerline{\normalsize DESY 97--036\hfill ISSN 0418--9833}
\centerline{\normalsize MPI/PhT/97--018\hfill}
\centerline{\normalsize hep--ph/9703280\hfill}
\centerline{\normalsize March 1997\hfill}}
\vskip1.5cm
Color-Octet Contributions to $J/\psi$ Photoproduction via Fragmentation at
HERA}
\author{{\sc Bernd A. Kniehl$^1$ and Gustav Kramer$^2$}\\
$^1$ Max-Planck-Institut f\"ur Physik (Werner-Heisenberg-Institut),\\
F\"ohringer Ring 6, 80805 Munich, Germany\\
$^2$ II. Institut f\"ur Theoretische Physik\thanks{Supported
by Bundesministerium f\"ur Forschung und Technologie, Bonn, Germany,
under Contract 05~7~HH~92P~(5),
and by EEC Program {\it Human Capital and Mobility} through Network
{\it Physics at High Energy Colliders} under Contract
CHRX--CT93--0357 (DG12 COMA).},
Universit\"at Hamburg,\\
Luruper Chaussee 149, 22761 Hamburg, Germany}
\date{}
\maketitle
\begin{abstract}
We study $J/\psi$ photoproduction via fragmentation at next-to-leading order
in the QCD-improved parton model, using the nonrelativistic factorization
formalism proposed by Bodwin, Braaten, and Lepage.
We consider direct and resolved photoproduction of prompt $J/\psi$ mesons and
$\chi_{cJ}$ mesons radiatively decaying to $J/\psi+\gamma$, taking into
account the formation of both color-singlet and color-octet $c\bar c$
states.
Adopting the values of the long-distance color-octet matrix elements
extracted from fits to prompt-$J/\psi$ data recently taken at the Fermilab
Tevatron, we predict that measurements of $J/\psi$ photoproduction at DESY
HERA should show a distinctive excess over the expectation based on the
color-singlet model at small values of the inelasticity variable $z$.
This is complementary to the expected enhancement at $z\alt1$ due to the
color-octet contribution to photon-gluon fusion.

\end{abstract}
\newpage

Some time ago, experiments \cite{abe} at the Fermilab $p\bar p$ collider
Tevatron have revealed that the production rate of prompt charmonium at large
transverse momentum ($p_T$) exceeds the most accurate theoretical predictions
within the so-called color-singlet model (CSM) \cite{csm} by more than one
order of magnitude.
This surprising observation may be interpreted by taking into account two
recent developments that have revolutionized the theoretical description of
heavy-quarkonium production in high-energy collisions.
Firstly, it has been realized \cite{bra} that fragmentation production must
dominate at large $p_T$, which implies that most charmonium in this kinematic
domain is produced by the hadronization of individual high-$p_T$ partons.
Secondly, a rigorous factorization formalism for quarkonium production based
on nonrelativistic quantum chromodynamics (QCD) has been developed \cite{bod},
which allows for a systematical treatment of the formation of charmonium from
color-singlet and color-octet $c\bar c$ pairs.
This formalism comprises a separation of short-distance parts, which are
amenable to perturbative QCD, from long-distance matrix elements, which must be 
computed through lattice simulations or extracted from experiment, and it takes 
into account the complete structure of the charmonium Fock space.
An important new feature of this approach is the appearance of the so-called
color-octet processes, in which a $c\bar c$ pair is produced at short
distances in a color-octet state and subsequently evolves into physical
(color-singlet) charmonium by the nonperturbative emission of soft gluons.
In fact, the Tevatron data \cite{abe} on prompt charmonium production at large
$p_T$ can be successfully interpreted by including the appropriate
color-octet processes and adjusting their long-distance matrix elements to
fit these data \cite{cho}.

In order to convincingly establish the phenomenological significance of the
color-octet mechanism, it is indispensable to identify the color-octet
contributions in other kinds of high-energy experiments as well.
On the theoretical side, a first step in this direction has recently been
taken in Ref.~\cite{che} by investigating the signature of color-octet
charmonium production in $e^+e^-$ annihilation on the $Z$-boson resonance.
Although the experimental data from CERN LEP \cite{abr} are in agreement with
the predicted signature, they do not exclude the hypothesis that the observed
prompt-charmonium signal is produced by color-singlet processes alone.
The analysis of low-energy $e^+e^-$ data might be more instructive \cite{yua}.

In this somewhat unsatisfactory situation, it would be very desirable if the
experiments at the DESY $ep$ collider HERA could probe the color-octet
mechanism of charmonium production and independently measure the magnitude of
the color-octet matrix elements.
HERA is presently operated in such a way that $E_e=27.5$~GeV positrons collide 
head-on with $E_p=820$~GeV protons in the laboratory frame, so that
approximately 300~GeV are available in the center-of-mass (CM) system.
Charmonium production at HERA dominantly proceeds via photoproduction, where
the incident positrons (or electrons) are scattered by small angles and
effectively act as a source of quasi-real photons, which interact with the
constituents of the incoming protons via hard scattering.
The scattering can either be elastic, diffractive, or inelastic.
In elastic scattering the protons stay intact, while in diffractive proton
dissociation they are destroyed by a small momentum transfer from the photons.
Both mechanisms may be interpreted by assuming that the protons interact with
their environs by the exchange of so-called pomerons, so that an adequate
theoretical treatment lies beyond the scope of the ordinary QCD-improved
parton model.
For a dedicated study of the color-octet processes, it is thus necessary to
focus attention on inelastic scattering.
In terms of the inelasticity variable
$z=p_p\cdot p_{J/\psi}/p_p\cdot p_\gamma$, where $p_p$, $p_\gamma$, and
$p_{J/\psi}$ are the proton, photon, and $J/\psi$ four-momenta, respectively,
this may be achieved by discarding events with $z$ values close to unity.
As may be inferred from the relation $z=2E_pm_T\exp(-y_{\rm lab})/W^2$, where
$y_{\rm lab}$ is the $J/\psi$ rapidity in the laboratory frame (with
$y_{\rm lab}>0$ in the proton flight direction),
$m_T=(M_{J/\psi}^2+p_T^2)^{1/2}$ is its transverse mass, and $W$ is the
$\gamma p$ CM energy, inelastic events tend to have large $y_{\rm lab}$ and
large $W$.

Recently, the H1 \cite{aid} and ZEUS \cite{col} collaborations presented
their 1994 data on the inelastic photoproduction of $J/\psi$ mesons, collected
in the kinematical ranges $0.45<z<0.9$, 30~GeV$<W<150$~GeV and $0.3<z<0.8$,
60~GeV$<W<130$~GeV, respectively.
After dividing out the respective integrated photon flux factors, these data
were compared \cite{aid,col} with a next-to-leading-order (NLO) calculation
\cite{kra} of $\gamma+p\to J/\psi+X$ via $\gamma g$ fusion \cite{ber} within
the CSM, where $W$ was fixed at the representative value $W=100$~GeV.
Although reasonable agreement was found \cite{aid,col}, one should bear in
mind that the theoretical prediction strongly depends on the chosen input 
values for the charm-quark mass $m_c$, the asymptotic scale parameter 
$\Lambda$, the renormalization scale $\mu$, and the factorization scale $M_f$,
with the normalization uncertainty being as large as a factor of three
\cite{kra}.
One might hence conclude that there is still some room left for alternative
$J/\psi$ production mechanisms.

Very recently, the analysis of Ref.~\cite{kra} was complemented \cite{cac} at
leading order (LO) by including the $2\to 2$ color-octet processes
$\gamma+g\to c\bar c[\,\underline{8},n]+g$ and
$\gamma+q\to c\bar c[\,\underline{8},n]+q$, where
$n={}^1\!S_0,{}^3\!S_1,{}^3\!P_J$ with $J=0,1,2$.
Adopting the color-octet matrix elements determined from fits \cite{cho} to
the Tevatron data \cite{abe} on prompt $J/\psi$ production, the authors of
Ref.~\cite{cac} found that the channels
$\gamma+g\to c\bar c[\,\underline{8},n]+g$ with
$n={}^1\!S_0,{}^3\!P_0,{}^3\!P_2$, the cross sections of which are divergent
in the limit $z\to1$, due to $g\to gg$ collinear splitting, dramatically
increase the cross section of inelastic $J/\psi$ photoproduction at HERA in
the upper $z$ range, at $z\agt0.8$.
On the other hand, the color-octet contribution rapidly falls off for $z$
decreasing, and is negligibly small already for $z\alt0.6$ \cite{cac}.

Being higher twist, $\gamma g$ fusion is only relevant if the produced
$J/\psi$ mesons have transverse momentum $p_T\alt M_{J/\psi}\approx3$~GeV.
In analogy to the situation at the Tevatron \cite{bra}, at
$p_T\gg M_{J/\psi}$, one expects fragmentation to be the dominant mechanism of
inelastic $J/\psi$ photoproduction at HERA, {\it i.e.}, single partons (mostly
$c$, $\bar c$ quarks and gluons) that are produced with high $p_T$ by the hard
$\gamma p$ scattering fragment into $J/\psi$ mesons.
The purpose of this Letter is to present a NLO analysis of this mechanism in
the framework of the QCD-improved parton model, and to assess the potential of
HERA to probe the color-octet processes in the range of low to intermediate
$z$ values, thus providing an independent check of the Tevatron measurements.

A first study \cite{god} of inelastic $J/\psi$ photoproduction via
fragmentation at HERA only included the Bethe-Heitler process
$\gamma+g\to c+\bar c$ and the Compton process $\gamma+q\to g+q$ followed by
$c\to J/\psi$ and $g\to J/\psi$ fragmentation at LO.
These processes contribute to direct photoproduction, where the quasi-real
photons couple directly to the quarks involved in the hard scattering.
In a fraction of time, the quasi-real photons fluctuate into bunches of quarks
and gluons, which in turn interact with the proton constituents via hard
scattering, while the photon remnants give rise to hadronic activity in the
backward direction (resolved photoproduction).
The longitudinal-momentum distributions of the partons inside the resolved
photons are described by parton density functions (PDF's).
At NLO, the direct- and resolved-photon contributions both strongly depend on
the factorization scheme and scale, and must be combined in order to give a 
well-defined physical observable.
From previous experience in connection with inclusive pion, kaon \cite{bin},
and $D^{*\pm}$ \cite{spi,jan} photoproduction at HERA, one may infer that, in 
the case of $J/\psi$ photoproduction via fragmentation, the resolved-photon
contribution, which was disregarded in Ref.~\cite{god}, is likely to be much
more important than the direct one.
Furthermore, one expects a substantial enhancement due to the NLO corrections,
which were neglected in Ref.~\cite{god}.

The factorization formalism developed in Ref.~\cite{bod} implies that the
charmonium fragmentation functions have the general form
\begin{equation}
D_{i\to H}(x,\mu)=\sum_nd_{i\to n}(x,\mu)\langle0|{\cal O}^H[n]|0\rangle,
\end{equation}
where $d_{i\to n}(x,\mu)$ gives the probability for the parton $i$ to form a
jet that includes a $c\bar c$ pair in the state labelled by $n$
carrying the longitudinal-momentum fraction $x$, and
$\langle0|{\cal O}^H[n]|0\rangle$ measures the probability for a pointlike
$c\bar c$ pair in the state $n$ to bind to form a physical charmonium state
$H$.
The coefficient $d_{i\to n}(x,\mu_0)$ at the initial scale
$\mu_0=2m_c\approx M_{J/\psi}$ involves only momenta of order $m_c$, and can
thus be calculated in nonrelativistic QCD as a perturbation expansion in the
running coupling constant $\alpha_s(\mu_0)$ \cite{bra}.
The evolution of $D_{i\to H}(x,\mu_0)$ up to higher fragmentation scales $\mu$
is ruled by the timelike Altarelli-Parisi equations, which may be conveniently
solved at NLO in $x$ space \cite{jan}.

We take into account prompt $J/\psi$ mesons as well as $\chi_{cJ}$ mesons
($J=0,1,2$) that radiatively decay to $J/\psi+\gamma$ with well-known
branching fractions \cite{pdg} (non-prompt $J/\psi$ mesons).
The dominant $J/\psi$ ($\chi_{cJ}$) Fock states are
$[\,\underline{1},{}^3\!S_1]$ and $[\,\underline{8},{}^3\!S_1]$
($[\,\underline{1},{}^3\!P_J]$ and $[\,\underline{8},{}^3\!S_1]$),
respectively.
The scaling of the respective matrix elements with the mass $m_c$ and the
relative velocity $v$ of the bound $c$ and $\bar c$ quarks is indicated in
Table~\ref{t1}.
In the case of $J/\psi$ ($\chi_{cJ}$), the color-singlet matrix element is
related to the (derivative of the) nonrelativisitic radial wave at the origin
and may thus be extracted from the measured leptonic (light hadronic)
annihilation rate \cite{bfy}.
The color-octet matrix elements have been determined from fits \cite{cho} to
Tevatron data \cite{abe}.
The input values \cite{bfy} adopted in our numerical analysis are also listed
in Table~\ref{t1}.
In the case of $D_{g\to J/\psi}$, the $v^4$ suppression of
$\langle0|{\cal O}^{J/\psi}[\,\underline{8},{}^3\!S_1]|0\rangle$ relative to
$\langle0|{\cal O}^{J/\psi}[\,\underline{1},{}^3\!S_1]|0\rangle$ is
numerically compensated by the fact that
$d_{g\to[\,\underline{8},{}^3\!S_1]}$ has two powers of $\alpha_s$ less than
$d_{g\to[\,\underline{1},{}^3\!S_1]}$.
Such a compensation does not occur for $D_{c\to J/\psi}$ and
$D_{c\to\chi_{cJ}}$, so that color-octet fragmentation is negligible in these
cases.
The color-singlet and color-octet contributions to $D_{g\to\chi_{cJ}}$ are 
theoretically intertwined, since $d_{g\to[\,\underline{1},{}^3\!P_J]}$ has a
logarithmic infrared singularity, which is absorbed into
$\langle0|{\cal O}^{\chi_{cJ}}[\,\underline{8},{}^3\!S_1]|0\rangle$.

We work at NLO in the $\overline{\rm MS}$ scheme with $n_f=4$ massless quark
flavors, $\Lambda_{\overline{\rm MS}}^{(4)}=296$~MeV \cite{lai}, and
$\mu=M_f=m_T$.
We take $c$ and $\bar c$ to be active partons inside the proton and resolved
photon, for which we adopt the CTEQ4M \cite{lai} and GRV \cite{grv} PDF's,
respectively.
As described in Ref.~\cite{spi}, we adjust the factorization of the
final-state collinear singularities associated with $c$ and $\bar c$ in such a
way that it matches the massive calculation.
We treat the quasi-real photon spectrum in the Weizs\"acker-Williams 
approximation with a maximum virtuality of $Q_{\rm max}^2=4$~GeV$^2$ as
described in Ref.~\cite{spi}.
We compare our fragmentation results with the LO prediction for direct
$J/\psi$ photoproduction via $\gamma g$ fusion within the CSM \cite{ber},
which we evaluate with the QCD-corrected value of
$\langle0|{\cal O}^{J/\psi}[\,\underline{1},{}^3\!S_1]|0\rangle$ specified in 
Table~\ref{t1}.
In this way, we include the bulk of the NLO enhancement, the residual
dynamical QCD correction being of order 20\% \cite{kra}.
Direct $\chi_{cJ}$ photoproduction via $\gamma g$ fusion is forbidden at LO in
the CMS, and it is marginal in the color-octet channel \cite{mca}.
Resolved $J/\psi$ photoproduction mediated by $gg$ fusion, which receives both
color-singlet and color-octet contributions, visibly contributes only at
very low $z$ and does not change the qualitative picture of $\gamma g$ fusion
\cite{mca}.

\begin{figure}[ht]
\begin{center}
\epsfxsize=16cm
\epsffile[54  172  541  614]{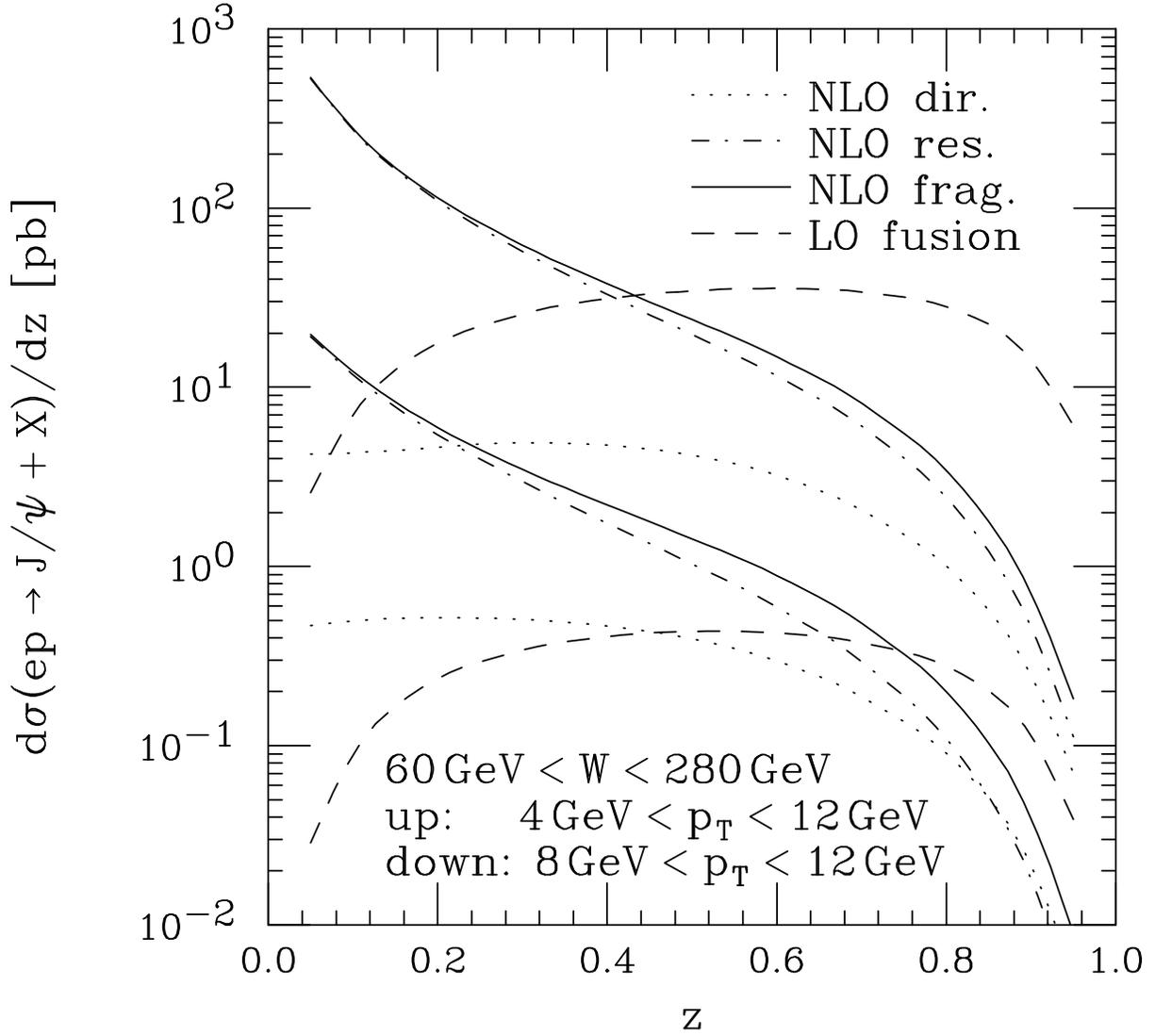}
\smallskip
\caption{Cross section $d\sigma/dz$ of inelastic $J/\psi$ photoproduction at
HERA, integrated over 60~GeV${}<W<280$~GeV and 4~GeV${}<p_T<12$~GeV
(upper curves) or 8~GeV${}<p_T<12$~GeV (lower curves).
The NLO fragmentation contributions due to direct photons (dotted lines),
resolved photons (dot-dashed lines), and their sum (solid lines) are compared
with the LO $\gamma g$-fusion contribution (dashed lines).}
\label{f1}
\end{center}
\end{figure}

\begin{figure}[ht]
\begin{center}
\epsfxsize=16cm
\epsffile[54  168  539  604]{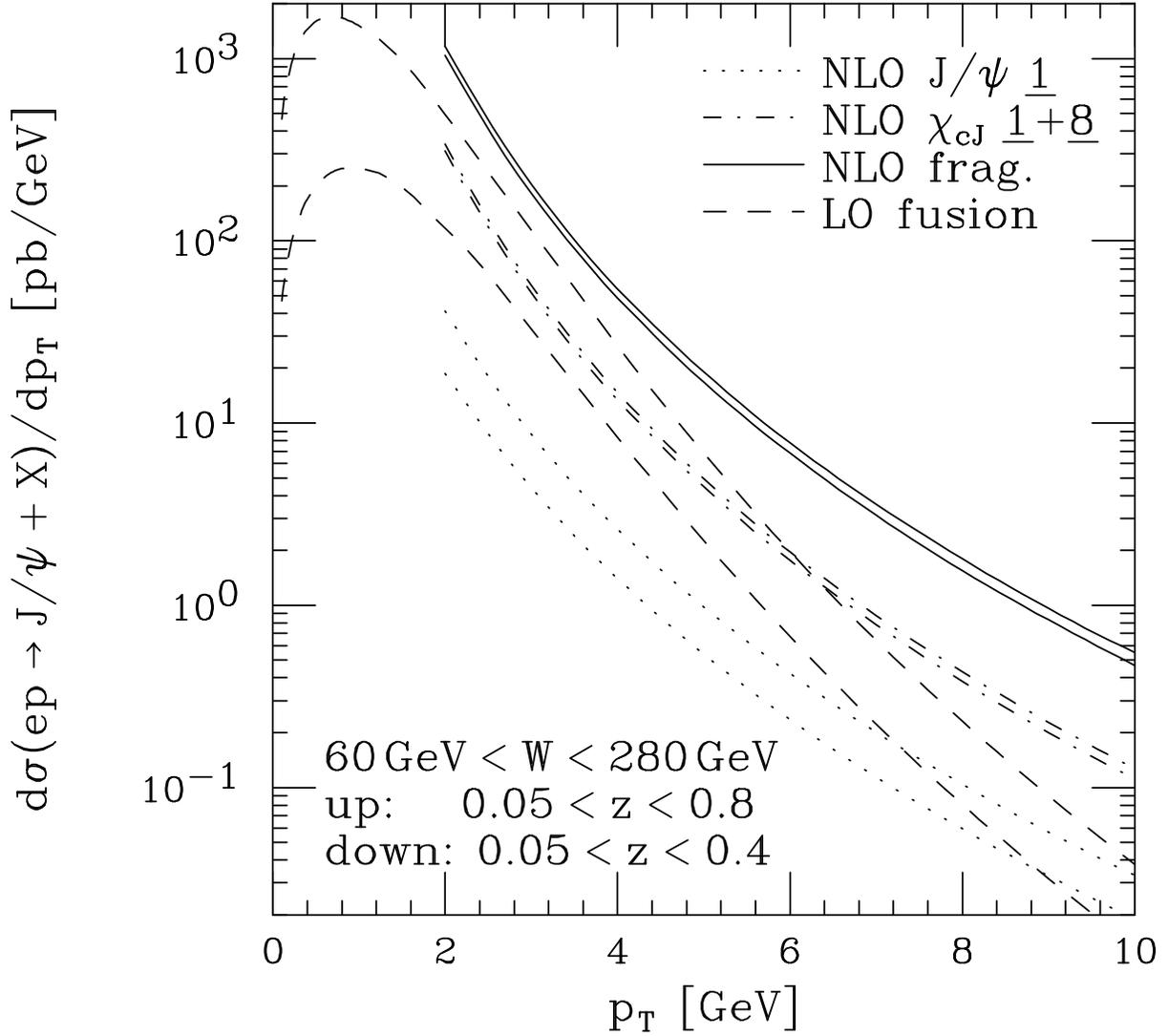}
\smallskip
\caption{Cross section $d\sigma/dp_T$ of inelastic $J/\psi$ photoproduction at
HERA, integrated over 60~GeV${}<W<280$~GeV and $0.05<z<0.8$ (upper curves) or
$0.05<z<0.4$ (lower curves).
The total NLO fragmentation contribution (solid lines) is compared with the LO
$\gamma g$-fusion contribution (dashed lines).
For comparison, also the prompt color-singlet (dotted lines) and the 
non-prompt (dot-dashed lines) fragmentation contributions are shown.}
\label{f2}
\end{center}
\end{figure}

\begin{figure}[ht]
\begin{center}
\epsfxsize=16cm
\epsffile[52  168  539  604]{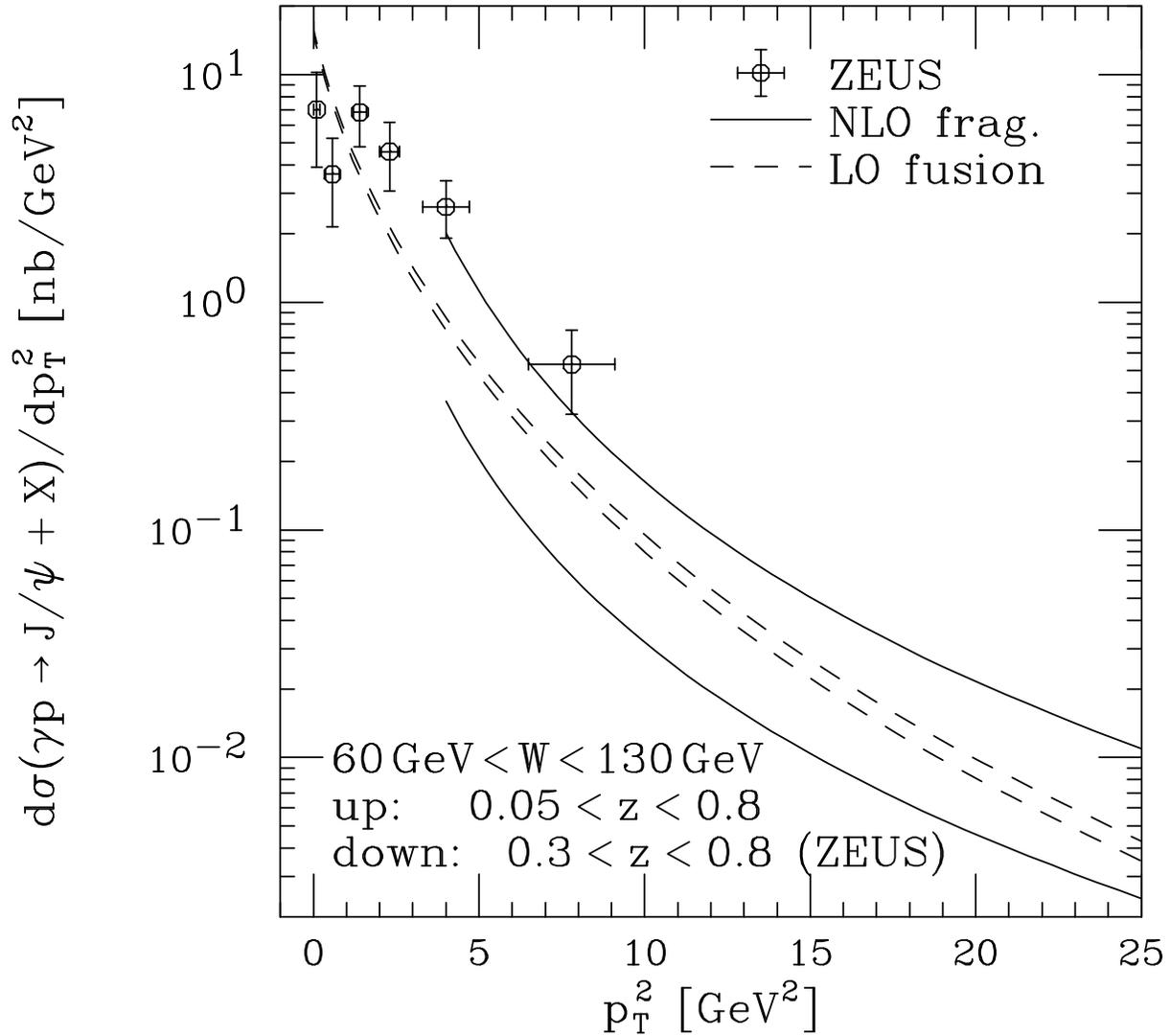}
\smallskip
\caption{Cross section $d\sigma/dp_T^2$ of inelastic $J/\psi$ photoproduction
at HERA, integrated over 60~GeV${}<W<130$~GeV and $0.3<z<0.8$ (lower curves)
and divided by the corresponding photon flux factor, $6.66\times10^{-2}$.
The total NLO fragmentation (solid lines) and LO $\gamma g$-fusion (dashed
lines) contributions are compared with the ZEUS data \protect\cite{col}.
For comparison, the theoretical results are also shown for $0.05<z<0.8$ (upper
curves).}
\label{f3}
\end{center}
\end{figure}

From Fig.~\ref{f1}, we observe that the fragmentation mechanism vastly
dominates inelastic $J/\psi$ photoproduction in the lower $z$ range.
For a minimum-$p_T$ cut of 4~GeV (8~GeV), its contribution exceeds the one due
to $\gamma g$ fusion for $z<0.4$ (0.75), by factors of about 4 and 200 (20 and
700) at $z=0.25$ and 0.05, respectively.
The bulk of the fragmentation contribution is induced by resolved photons,
except for $z$ close to unity, where fragmentation is anyway insignificant.
Since we wish to elaborate color-octet signatures at low and intermediate
$z$, we regard $\gamma g$ fusion, which for $z\alt0.6$ is well described by
the CSM \cite{cac}, as a background.
As is evident from Fig.~\ref{f2}, we may efficiently suppress this background
by introducing a stringent maximum-$z$ cut, of 0.4 say.
In fact, if the maximum-$z$ cut is lowered from 0.8 to 0.4, the cross section
of $\gamma g$ fusion at $p_T=2$~GeV (10~GeV) is reduced by a factor of 4.2
(2.8), while that of fragmentation is only insignificantly decreased, by 11\%
(15\%).
At $p_T=2$~GeV (10~GeV), 2\% and 30\% (4\% and 24\%) of the fragmentation
cross section integrated over $0.05<z<0.4$ are due to the prompt color-singlet
and non-prompt channels, respectively.
Thus, this regime offers a unique laboratory to probe the prompt color-octet
channel and thus to measure
$\langle0|{\cal O}^{J/\psi}[\,\underline{8},{}^3\!S_1]|0\rangle$, so as to
check the Tevatron result.
Even for $0.05<z<0.8$, the fragmentation cross section in Fig.~\ref{f2} is in
excess of the one due to $\gamma g$ fusion.
As may be inferred from Fig.~\ref{f1}, this is subject to change if the
minimum-$z$ cut is increased.
This is illustrated in Fig.~\ref{f3}, where the ZEUS $J/\psi\to\mu^+\mu^-$
data \cite{col} collected in the interval $0.3<z<0.8$ are compared with the
respective fragmentation and $\gamma g$-fusion predictions.
We have divided the experimental data points by the estimated extrapolation
factor 1.07 which was included in Ref.~\cite{col} to account for the
unmeasured contribution from $0<z<0.3$.
In fact, our combined analysis of fragmentation and $\gamma g$ fusion suggests
that, at $p_T=2$~GeV (5~GeV), this factor should be as large as 5.2 (3.4).
The corresponding value for $0.05<z<0.3$ is 2.6 (2.6).
The gap between the combined result for $0.3<z<0.8$ and the ZEUS data is
partly filled by the dynamical NLO corrections \cite{kra} and the color-octet
contributions \cite{cac} to $\gamma g$-fusion, which are not included in
Fig.~\ref{f3}.

In conclusion, the cross section of inelastic $J/\psi$ photoproduction in $ep$ 
collisions at low $z$ and large $p_T$ is very sensitive to the color-octet
matrix element
$\langle0|{\cal O}^{J/\psi}[\,\underline{8},{}^3\!S_1]|0\rangle$.
We propose to accordingly extend previous measurements of this cross section
at HERA, in order to obtain an independent, nontrivial check of the Tevatron
color-octet charmonium puzzle.

\begin{table}
\begin{center}
\caption{Values \protect\cite{bfy} of the matrix elements used in the
numerical analysis and their mass and velocity scaling.
The relations between the $J$-dependent matrix elements follow from 
heavy-quark spin symmetry.}
\label{t1}
\smallskip
\begin{tabular}{lrl}
\hline
\hline
$\langle0|{\cal O}^{J/\psi}[\,\underline{1},{}^3\!S_1]|0\rangle$ &
1.13~GeV$^3$ & $[m_c^3v^3]$ \\
$\langle0|{\cal O}^{J/\psi}[\,\underline{8},{}^3\!S_1]|0\rangle$ &
0.014~GeV$^3$ & $[m_c^3v^7]$ \\
$\langle0|{\cal O}^{\chi_{cJ}}[\,\underline{1},{}^3\!P_J]|0\rangle/(2J+1)$ &
0.0880~GeV$^5$ & $[m_c^5v^5]$ \\
$\langle0|{\cal O}^{\chi_{cJ}}[\,\underline{8},{}^3\!S_1]|0\rangle/(2J+1)$ &
0.0076~GeV$^3$ & $[m_c^3v^5]$ \\
\hline
\hline
\end{tabular}
\end{center}
\end{table}

\end{document}